\documentclass[a4paper]{article}

\usepackage{INTERSPEECH2022}
\usepackage{amsmath,graphicx}
\usepackage{textcomp}
\usepackage{microtype}
\usepackage{booktabs}
\usepackage{multirow}
\usepackage{cite}
\usepackage{placeins}
\usepackage{float}
\usepackage{balance}
\usepackage[ruled,noend]{algorithm2e} 

\title{Reducing Geographic Disparities in Automatic Speech Recognition \\ via Elastic Weight Consolidation}
\name{Viet Anh Trinh, Pegah Ghahremani, Brian King, \\ Jasha Droppo, Andreas Stolcke and Roland Maas}
\address{Amazon Alexa AI, USA}
\email{\{trinhvie,pegahgh,bbking,drojasha,stolcke,rmaas\}@amazon.com}

\begin{document}

\maketitle
\begin{abstract}

We present an approach to reduce the performance disparity between geographic regions without degrading performance on the overall user population for ASR. A popular approach is to fine-tune the model with data from regions where the ASR model has a higher word error rate (WER). However, when the ASR model is adapted to get better performance on these high-WER regions, its parameters wander from the previous optimal values, which can lead to worse performance in other regions.
In our proposed method, we utilize the elastic weight consolidation (EWC) regularization loss to identify directions in parameters space along which the ASR weights can vary to improve for high-error regions, while still maintaining performance on the speaker population overall. Our results demonstrate that EWC can reduce the word error rate (WER) in the region with highest WER by 3.2\% relative while reducing the overall WER by 1.3\% relative. We also evaluate the role of language and acoustic models in ASR fairness and propose a clustering algorithm to identify WER disparities based on geographic region. 
\end{abstract}
\section{Introduction}
\label{sec:intro}
Automatic speech recognition (ASR) has achieved high accuracy by adopting a range of neural network architectures \cite{chan2016listen, li2021recent,  DBLP:journals/corr/abs-1211-3711, he2019streaming, guo2020efficient, zhang2020transformer,watanabe2018espnet}. One remaining challenge is that ASR performance can vary substantially by speaker~\cite{tatman2017effects, koenecke2020racial,martin2020understanding,tatman2017gender, gao2022seamless, sari2021counterfactually,liu2021model}. The problem is associated with a variety of factors, including geographic location, age, and accent. In this research, we focus on reducing disparity in ASR performance between different geographic regions.

We can enhance the performance for an underperforming region by collecting more training data from that region. In general, there are two mechanisms to employ this additional data: retraining the ASR model from scratch with both current data and the new data from high-error regions, or adapting the already-trained model with data from the high-error regions. The first mechanism can be computationally expensive and requires access to past data, which is not always possible. However, it might lead to better performance than the second approach. The second mechanism can save training time, and is suitable for many real-world scenarios, such as adapting the ASR model on edge devices \cite{peinl2020open} with local data, where access to the cloud-based training data is impractical. Here we focus on the second mechanism (adaptation) and make use of the first mechanism (retraining) as an empirical upper bound on performance. 

In this paper, we employ elastic weight consolidation (EWC)~\cite{kirkpatrick2017overcoming} to adapt the pretrained ASR model. This method is designed to keep the model from forgetting important parameters learned in the initial training while acquiring new knowledge. More specifically, EWC adds a regularization term to the ASR loss to make the model parameters of the new task (recognition of regions having high WER) to be close to the best parameters for the previous task (recognition of all regions), along the dimensions that matter the most to the previous task. We compare the EWC method with several transfer learning \cite{pan2009survey} techniques.

EWC belongs to the continual-learning \cite{delange2021continual} family of methods, referring to the ability of a machine learning model to learn continually from new data being ingested over time \cite{parisi2019continual}. Regularization and memorization-based techniques are two popular directions to deal with the forgetting problem arising in continual learning. 

In the regularization-based approach, there are ``importance scores'' assigned to the parameter weights, so the model has the flexibility to move along the dimensions that are not important for the old task. Generally, the regularization-based approach has a loss of the form
\begin{equation}
\mathcal{L}(\theta)= \mathcal{L}_{new}(\theta) + \lambda \sum_{k}  I_k (\theta_{k} - \theta_{old,k}^*)^2
\end{equation}
where $\mathcal{L}$ is the loss, $\theta_{k}$ is the $k$th parameter for the new task, $\theta_{old,k}^*$ is the same parameter for the old task at convergence, $I_k$ is the importance score for the $k$th parameter, and the scalar $\lambda$  controls the balance between learning the new task and remembering the old task. There are several ways to assign the importance scores. While EWC utilizes an off-line Fisher information matrix, synaptic intelligence (SI) \cite{zenke2017continual} computes the importance scores online, and memory-aware synapse (MAS) \cite{aljundi2018memory} uses the squared (L2) norm of the gradient of the output with respect to the parameters.

In other related work, \cite{riemer2018learning} proposes the meta-experience relay (MER) method, which utilizes a fixed-size memory buffer holding random samples from both old and new tasks, and optimizes the network within the meta-learning framework. Additionally, \cite{nguyen2018variational} presents the variational continual learning (VCL) method, which uses a Bayesian core set (similar to memory) sampled by K-center \cite{gonzalez1985clustering} or randomly. In \cite{nguyen2018variational}, the old task is the prior and the new task is the posterior in the Bayesian framework, and Bayesian variational inference is applied to approximate the posterior distribution. 

In this paper, we propose a loss function to which an additional EWC regularization term is added to make the model perform better in high-WER regions, without forgetting knowledge about the general user population. In Section~\ref{sec:methods} we describe our method in detail.  Section~\ref{sec:experiments} describes the data, experiments and training parameters. Results are presented in Section~\ref{sec:results}. 

\section{Methods}
    \label{sec:methods}
    
\subsection{Problem statement}
 As input to our task, we are given an ASR model  pretrained on a dataset of speech from the general user population. This pretrained model has good performance on average, but has high error rate for speakers from some geographic areas.  Our task is to eliminate (or reduce) the performance gap \cite{mehrabi2021survey} against these regions, without access to the data of the pretraining stage and without degrading average performance for all regions. 

\subsection{ASR Model}
Our ASR system is an end-to-end model utilizing the recurrent neural network transducer (RNN-T) architecture \cite{DBLP:journals/corr/abs-1211-3711, he2019streaming}, which is suitable for streaming. More specifically, the system includes a long short-term memory (LSTM) \cite{hochreiter1997long} encoder, an LSTM prediction network, and a joint network. At each time frame, the encoder output is a vector $\mathbf{h}_t^{enc}$ that summarizes input information over time:
\begin{equation}
    \mathbf{h}_t^{enc} = \mathrm{LSTM}(\mathbf{x}_{t})
 \end{equation}
where $\mathbf{x}_{t}$ is the vector of acoustic features at time $t$. 
The predictor takes the previous predicted labels $\mathbf{\hat{y}}_{m-1}$ as the input and returns $\mathbf{h}_m^{pred}$ as the output:
\begin{equation}
    \mathbf{h}_m^{pred} = \mathrm{LSTM}(\mathbf{\hat{y}}_{m-1}) \\
\end{equation}
The outputs of the encoder and predictor are combined by a joint network, which is a simple feed-forward network (FFN) \cite{li2019improving}:
\begin{equation}
    F^{joint}(\mathbf{h}_t^{enc},\mathbf{h}_m^{pred}) = \mathrm{tanh}(\mathbf{W}\mathbf{h}_t^{enc} + \mathbf{V}\mathbf{h}_m^{pred})
\end{equation}
where $\mathbf{W}$ and $\mathbf{V}$ are weight matrices.
Finally, a softmax layer is applied to the output of the joint network to make the final prediction:
\begin{equation}
    p(\mathbf{\hat{y}}_{m}|\mathbf{x}_{1:t},\mathbf{\hat{y}}_{1:m-1}) = \mathrm{softmax}(F^{joint}(\mathbf{h}_t^{enc},\mathbf{h}_m^{pred}))
\end{equation}

\subsection{Elastic weight consolidation}

Elastic weight consolidation (EWC) \cite{kirkpatrick2017overcoming} can be applied to make the model acquire new knowledge while retaining information already learned. We denote model parameters by $\mathbf{\theta}$, and old and new data by $D_A$ and $D_B$, respectively.
Applying Bayes' theorem
\begin{align} \label{eq:1}
    \log p(\mathbf{\theta}|D_A, D_B) &= \log p(D_B|\mathbf{\theta}) + \log p(\mathbf{\theta} | D_A) \nonumber \\
    &- \log p(D_B | D_A)
\end{align}

EWC approximates $\log p(\mathbf{\theta} |D_A)$ with the second Taylor expansion at $\theta_A^*$. As $\log p'(\mathbf{\theta}_A^*|D_A) = 0$  at $\mathbf{\theta}_A^*$,
\begin{align}
    \log p(\theta|D_A) &\approx \log p(\theta_A^*|D_A) + \frac{1}{2} \log p'' (\theta_A^* | D_A)(\theta - \theta_A^*)^2 \nonumber \\
    & \approx const + \frac{1}{2} \log p'' (\theta_A^* | D_A)(\theta - \theta_A^*)^2
\end{align}
We observe that a normal distribution $\mathcal{N}(\mu,\sigma^2)$ has log probability density in the form $constant - \frac{1}{2\sigma^2}(x-\mu)^2$, then
\begin{equation}
\log p(\theta|D_A) \approx \mathcal{N}(\theta_A^*,F^{-1})
\end{equation}
where $F = \frac{1}{2}\log p'' (\mathbf{\theta}_A^*|D_A)$ is the Fisher information matrix.
EWC assumes the Fisher information matrix is diagonal, giving
\begin{equation} \label{eq:2}
    \log p(\mathbf{\theta}|D_A)  \approx constant + \frac{1}{2} F_i (\mathbf{\theta} - \mathbf{\theta}_A^*)^2
\end{equation}
Because $\log p(D_B | D_A)$ is a constant, substituting $\log p(\mathbf{\theta}|D_A)$ from Eq.~(\ref{eq:2}) in Eq.~(\ref{eq:1}), we obtain
\begin{align} \label{eq:3}
    \log p(\mathbf{\theta}|D_AD_B) \approx \log p(D_B|\mathbf{\theta}) + \frac{1}{2} F_i (\mathbf{\theta} - \mathbf{\theta}_A^*)^2 
\end{align}

As we can see, there is a regularization term $\frac{1}{2} F_i (\mathbf{\theta} - \mathbf{\theta}_A^*)^2$ added to the model's loss on the new task $\log p(D_B|\mathbf{\theta})$, to keep the parameters on the new task  $\mathbf{\theta}_B$ close to the best parameters of the old task  $\mathbf{\theta}_A^*$ along the dimensions that matter to task A. $F_i$ can be interpreted as the importance scores that EWC assigns to the $i$th parameter, where $F_i$ is the element at position $i$ on the main diagonal of the Fisher information matrix.

\subsection{Loss function for adaptation}

Here, we formulate the problem of eliminating geographic disparities in a continual learning setup for ASR, where we adapt a pretrained model (trained on data from all regions) on new data that contains only speech from regions having high WER, without access to previous data. We propose a new loss function that is a combination of the EWC loss and the RNN-T loss. This loss function can mitigate the problem of distribution-shift between the pretraining data and fine-tuning (adaptation) data. 

Our loss function is defined as follows:
\begin{equation}
\mathcal{L}(\mathbf{\theta}) = \mathcal{L}_{ASR}(\mathbf{\theta}) + \frac{\lambda}{2}\sum_{i} F_i (\mathbf{\theta}_{i} - \mathbf{\theta}_{p,i}^*)^2 
\label{eq:4}
\end{equation}
As before, the additional EWC regularization term forces the parameters of the ASR model trained on high-WER regions $\mathbf{\theta}$ to be close to the best parameters of the model trained on all regions  $\mathbf{\theta}_p^*$, along the dimensions that are important to the pretrained task. More specifically, every parameter of the ASR model has a different penalty when it moves away from the optimal pretrained value. The parameters that are important to the pretraining task have high penalties, so they cannot change much from the prior optimal values. On the other hand, the parameters that are less essential to the pretraining task are allowed to vary more freely. Thus, the ASR model can still improve performance during fine-tuning for regions with high WER.  

Our work follows \cite{kunstner2019limitations} in approximating the Fisher information matrix, where the empirical $F_i$ is derived as:
\begin{equation} \label{eq:fim}
    F_i = \sum_{j \in D} \mathbb{E}(\frac{\partial \mathcal{L}_{ASR}(y_j,\hat{y_j})}{\partial \theta_i})^2
\end{equation}
where $y_j$, $\hat{y}_j$ are the label and ASR output respectively, $D$ is the dataset used to extract the Fisher information matrix, $\mathcal{L}_{ASR}$ is the ASR loss, and $\theta_i$ is a parameter of the ASR model.  $F_i$ is the $i$th element on the main diagonal of the Fisher information matrix, which acts as an importance score for parameter $\theta_i$ with respect to the previous task, in our case reflecting how important parameter $\theta_i$ is to ASR performance on the general user population. 

\subsection{Identifying regions with high word error rate by geo-clustering}
\label{sec:tree}

To improve ASR performance for regions with high WER, we aim to adapt the model with additional data from those areas. Thus, we built a clustering tree to select training samples belonging to those regions. The clustering tree is trained and tested on a dataset separate from the ASR training set. The tree training data is split into different regions by using an algorithm that maximizes the WER differences between regions. The geographic clustering tree keeps splitting the data into left and right branches by either approximate longitude or latitude while the number of devices (as a proxy for users) in each leaf is larger than or equal to a predefined threshold. The decision to split by longitude or latitude depends on which coordinate yields the larger WER difference between the two  branches. Pseudo-code for geographic clustering is given in Algorithms~\ref{alg:alg1} and~\ref{alg:alg2}.

\textbf{
\SetAlgoNoLine
\DontPrintSemicolon
\LinesNumbered
{
\begin{algorithm}[t]
\scriptsize 
\SetKwFunction{bestsplit}{Best split}
\SetKwProg{Fn}{Function}{:}{}
\Fn{\bestsplit{data X, threshold t}}{
Best coordinate = None, WER difference = 0 \;
 \For{coordinate in \{longitude,latitude\}}
 {
$X_{left}$ = \{$x \in X|x <$ median coordinate \} \;
$X_{right}$ = \{$x \in X|x \geq$ median coordinate\} \;
  \If{$|X_{left}| < t$ or $|X_{right}| < t$ devices } {
   go to line 14  \;
   }
  \# $X_{left}$ WER is the WER of a dataset contains all \;
  \# utterances on the left branch \;
   d = ($X_{left}$ WER - $X_{right}$ WER)$^2$ \;
    \If{d $>$ difference}{
    Best coordinate = coordinate \;
    WER difference = d \;
    }
    \KwRet Best coordinate, WER difference}
}
\caption{Best split}
\label{alg:alg1}
\end{algorithm}}%
}

\SetAlgoNoLine
\DontPrintSemicolon
\LinesNumbered
{
\begin{algorithm}[t]
\scriptsize 
\SetKwFunction{bestsplit}{Best split}
\SetKwProg{Fn}{Function}{:}{}

\SetKwFunction{buildtree}{Build tree}
  \SetKwProg{Fn}{Function}{:}{}
  \Fn{\buildtree{data X, threshold t}}{
       Best-coordinate, WER-diff = \bestsplit{X, t} \;
        \If{WER-diff = 0 or (WER-diff $\neq$ 0 and Best-coord is None)}{
create new region; compute WER for new region \; }
	$X_{left}$ = \{$x \in X|x <$ median coordinate \} \;
$X_{right}$ = \{$x \in X|x \geq$ median coordinate\} \;
	$tree_{left}$ = {\buildtree{$X_{left}$, t}} \;
$tree_{right}$={\buildtree{$X_{right}$, t}} \;
\# tree is an object containing information about left, right tree \;
\# branches and the coordinate \;
        \KwRet tree \;
  }

 \caption{Geo-clustering tree}
 \label{alg:alg2}
\end{algorithm}}%


\begin{table*}[tb]
  \footnotesize
  \caption{Relative WER and variance reduction on the test set. Region WER for a dataset containing only utterances from a specific geographic region. Overall WER is over the whole test set. For the variance column, negative numbers indicate a smaller variance; for the other columns, a negative number indicates a relative WER improvement. $D_p$ is the pretraining dataset. $D_c$ and $D_r$ are two fine-tuning datasets, where $D_c$ contains utterances from the regions with high WER while $D_r$ has equally many randomly selected utterances. Each of $D_p$, $D_c$ and $D_r$ contain 10k hours of speech. WERR is WER reduction, lower is better.}
  \label{tbl:result1}
  \centering
  \begin{tabular}{llcrrrrr}
    \toprule
     \textbf{Experiment}   & \textbf{Description} & \textbf{Data} & \multicolumn{4}{c}{\textbf{Region WERR}}                          &  \textbf{Overall} \\
            \cline{4-7}
                           &                      &               & \textbf{variance} & \textbf{mean} & \textbf{max} & \textbf{min} &  \textbf{WERR} \\
                        \midrule
    Experiment 1 & Baseline & $D_p$ & 0	& 0 & 0 &	0	& 0 \\
    Experiment 2 & No freeze & $D_c$   & -5.3  & -0.9 &	-2.9	& -4.6 & -1.1 \\
    Experiment 3 & Freeze Encoder & $D_c$  & -1.8 & 0.0 & -1.4 & -5.4  & -0.1 \\
    Experiment 4 & Freeze Predictor & $D_c$  & 1.8 & -0.3& -1.3 & -8.5 & -0.4 \\
    Experiment 5 & Freeze 3 lowest encoder \newline layers and 1 predictor layer & $D_c$  & -0.9 & -0.5 & -2.5 & -2.7 & -0.4 \\
    Experiment 6 & Proposed method & $D_c$ & \textbf{-7.9} &	\textbf{-1.1} &	\textbf{-3.2}	& \textbf{-5.8} &	\textbf{-1.3} \\
    \midrule
    Experiment 7 & Empirical bound & $D_p$ +$D_c$& -5.3 & -1.2 & -2.3 & 0.2 & -1.0 \\
 Experiment 8 &  & $D_p$ +$D_r$& -12.3 & -2.3 & -0.9 & -7.3 & -2.1 \\
    \bottomrule
  \end{tabular}
\end{table*}

\section{Data and Experiments}
    \label{sec:experiments}

\subsection{Identifying geographic regions}
\begin{figure}[t]
\centering
\includegraphics[width=\columnwidth]{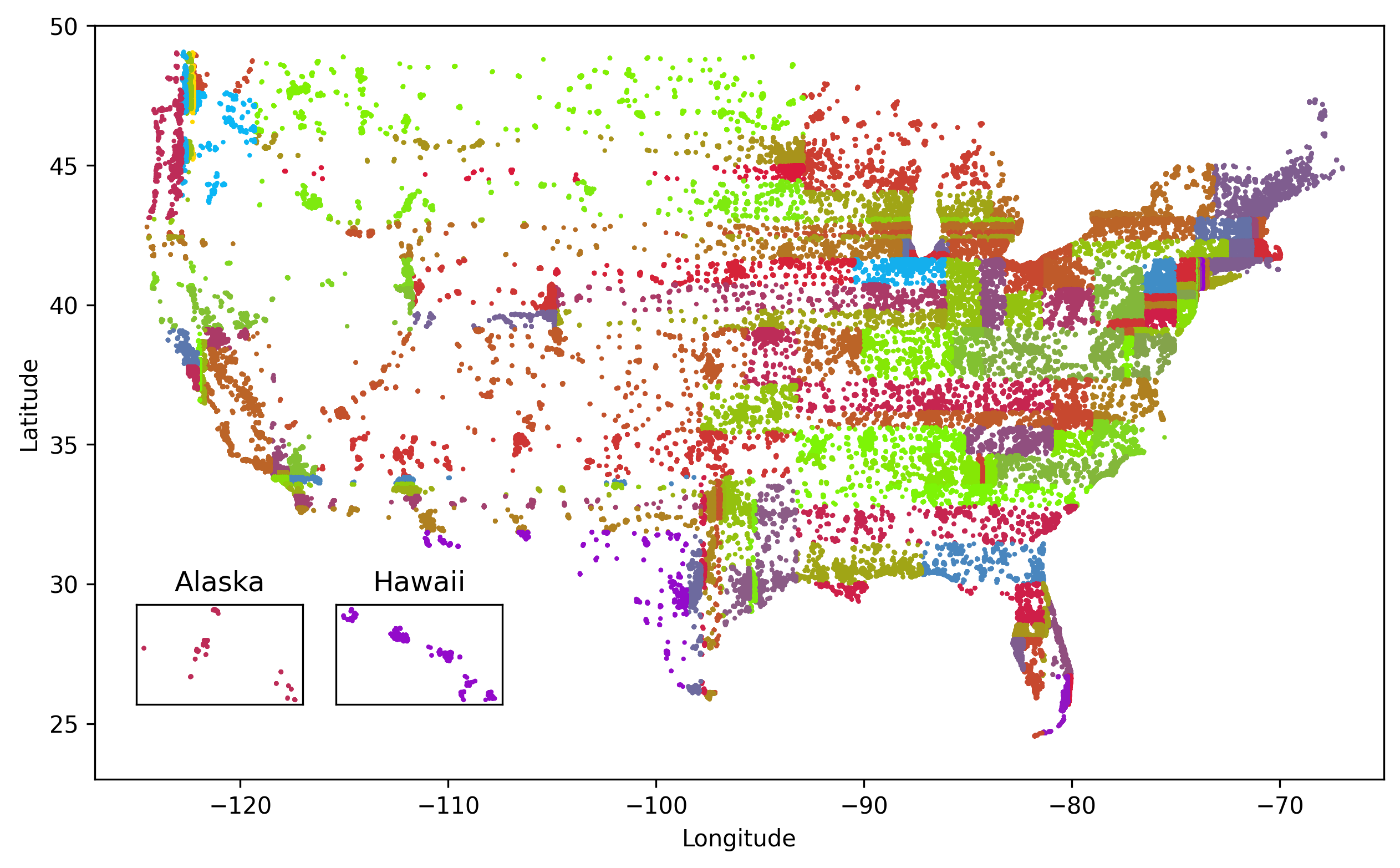}
\caption{126 regions identified by the clustering tree. The color does not indicate specific WER, however regions with the same color have the same WER.}
\label{fig:fig1}
\end{figure}

We utilize the clustering tree algorithm in Section~\ref{sec:tree} to split the dataset into different subsets, where each subset corresponds to a specific geographic region. Note that the data is not split by state, city or zip code, but by approximate longitude and latitude. The clustering tree is trained with 5-fold cross-validation on de-identified user data from a commercial voice-enabled artificial intelligence assistant, consisting of 1.1k hours of audio.  Thus for every fold, the training and test sets contain 0.9k and 0.2k hours, respectively. After 5-fold cross-validation, we obtain five different clustering trees. We select the tree with the lowest L1-distance between the predicted and true WERs per region, over all regions from the test set. The tree keeps splitting the dataset into smaller subsets, with the condition that each subset have at least $t$ devices. We evaluated different values of $t$ (1500, 2000 and 2500) and chose $t = 1500$, as this resulted in the largest WER disparity between highest and lowest region-specific WERs. The resulting tree split the dataset into 126 subsets, corresponding to 126 longitude/latitude-bounded geographic regions, as illustrated in Figure \ref{fig:fig1}. 

\subsection{Datasets}

In addition to the set of 1.1k hours for training the clustering tree, we used another de-identified dataset from the same system, comprising  47k hours of audio, to train the ASR system. In the transfer learning setup, there is a pretraining and a fine-tuning stage. As every stage requires a different dataset, we split the 47k-hour corpus ($D_w$) into several subsets. The ASR pretraining set ($D_p$) is created by drawing randomly 10k hours of speech from $D_w$. After removing the pretraining set, we use 37k hours of speech to form two fine-tuning sets. The first set $D_c$ is created by taking utterances from the regions in order of decreasing WER until $D_c$ contains 10k hours of speech. As a result, we ended up with 35 geographic regions in $D_c$. As a control, we sample 10k hours randomly from the 37k-hour set to form another fine-tuning set $D_r$. We found that $D_r$ and $D_c$ have 2.8k hours in common. The development and test sets comprise 25 and 125 hours, respectively. 

\subsection{Training parameters}
The encoder is a five-layer LSTM network with 1024 hidden units per layer. The predictor is a two-layer LSTM with 1024 units per layer. The joint network is a single-layer feedforward network with 512 hidden units. All these models are trained with Adam \cite{Adam} optimizer. The baseline is trained with a learning rate of $6.25\times 10^{-5}$. Other models are fine-tuned with an initial learning rate of $6.25\times 10^{-5}$, and a smaller learning rate of $1\times 10^{-5}$
 after 100k steps (every step uses 5 hours of speech). The $\lambda$ value in Eq.~(\ref{eq:4}) that defines the weight between the EWC regularization loss and the ASR loss is set to 1, giving ASR loss and EWC regularization equal influence. Given limited time and the large size of our training set, we did not optimize $\lambda$. The speech feature consists of 64-dimensional log-filterbank energies \cite{raeesy2018lstm}; sampling rate is 16\,kHz. Training uses Tensorflow. 

\subsection{Experiments}

We compare our proposed method against other transfer learning techniques. In Exp.~1 we train a baseline ASR model on the $D_p$ subset of 10k hours of random speakers. This baseline model after pretraining is used to initialize the weights of all other transfer learning methods (Exps.~2-5) and our proposed technique (Exp.~6). In Exp.~2, we adapt the pretrained model on the dataset $D_c$ without freezing any parameters. The encoder parameters are frozen in Exp.~3 while the predictor parameters are frozen in Exp.~4. In Exp.~5, the parameters of the first three layers of the encoders and the first layer of the predictor are frozen. The first layers are frozen because the representations at these early stages of the network often capture basic patterns of speech, which might transfer better to a new dataset than the representations at later layers, which we expect to capture more abstract information. 

In Exp.~6, we implement our proposed method with EWC regularization for RNN-Ts. In this experiment, the Fisher matrix is derived with the converged pretrained model from Exp.~1 following Eq.~(\ref{eq:fim}).
We then carry out Exp.~7 by training an ASR model with both pretraining $D_p$ and fine-tuning $D_c$ datasets from scratch, rather than adapting a pretrained model on $D_c$ as in Exps.~2-6. In  Exp.~7, there is no distribution shift because the training utterances are randomly selected from the union of $D_p$ and $D_c$. Exp.~7 represents the ideal case where we have all the data available for training, compared to the case of no access to pretraining data. In Exp.~8, we train the ASR from scratch with a dataset that is a combination of $D_p$ and $D_r$.

\section{Results and Discussion}
    \label{sec:results}

Results are shown in Table \ref{tbl:result1}. We report the relative test set error rate reductions compared to the baseline (which thus has relative WER 0).  We define a region WER as the WER of a dataset that contains only utterances belonging to a specific geographic region. In this way, we have 126 WER values for 126 geographic regions. We report the mean, variance, minimum and maximum of those 126 WER values. The WER over the whole test set (column 'Overall WER') is also reported.  

The results show that among the conventional transfer learning approaches (Exps.~2-5), the best result overall is obtained when the ASR parameters are not frozen (Exp.~2). In that case, the region WER max (from the region with the highest WER) is reduced by 2.9\% compared to the baseline, while freezing the encoder, or only its first three layers and the first predictor layer, reduces the region WER max by 1.4\% and 2.5\%, respectively.
Freezing the predictor (Exp.~4) reduces the region WER max by 1.3\%. Since the predictor predicts the next character given previous characters, it acts as a lightweight language model for the RNN-T. Therefore, freezing the predictor prevents the model from learning new linguistic information. Freezing the encoder (Exp.~3), on the other hand, restricts the ASR system from capturing new acoustic knowledge. The variance is reduced by 1.8\% in Exp.~3, while increasing by 1.8\% in Exp.~4; adapting the predictor (Exp.~3) is especially important for reducing variance between regions, indicating that language usage (words, grammar) contributes much to the performance disparity.

Moreover, our proposed method improves the geographic WER disparity the most, reducing both the WER of the region with highest WER (by 3.2\%) and the variance across regions (by 7.9\%), more than any other adaptation method investigated here. Also, the proposed approach reduces the WER on the whole test set by 1.3\%, versus the empirical bound of 1.0\% given by Exp.~7. Exp.~7 can be considered a notional optimum for adaptation because the model is trained on both pretraining dataset $D_p$ and fine-tuning dataset  $D_c$ jointly, while in the continual learning setup, only fine-tuning data may be used. In Exp.~8, the WER of the region with highest WER only decreases by $0.9\%$, as a result of using random data from $D_r$ compared to $3.2\%$ in Exp.~6 ($D_r$ has 2.8k hours in common with $D_c$). 

Our proposed method improves performance on the whole user population by 1.3\%, so is effective at preserving knowledge learned by the model from past data in the pretraining stage. More specifically, the proposed method can enhance performance in regions with the highest WER without sacrificing performance in other regions. Also, other conventional transfer-learning methods forget little past knowledge; e.g., Exp.~4 has the same overall WER performance as the baseline. We can credit the small learning rate in fine-tuning for this result. However, while a small learning rate mitigates the forgetting problem, it also limits their model's ability to learn new knowledge. Furthermore, the pretraining data already contains some of the regions with high WER, further alleviating forgetting.
Finally, although we have not explored the hyperparameter $\lambda$ in Eq.~(\ref{eq:4}), it could be used to  balance the learning of new knowledge against the forgetting of old knowledge in an operational setting. 

\section{Conclusion}

We have investigated a method to address geographic disparities and fairness in ASR \cite{mehrabi2021survey}. We propose an RNN-T loss function combining the standard ASR loss with EWC regularization loss to encourage the ASR model to find parameters that have good performance for the user population overall, while reducing the performance degradation for speakers from regions with high WER. Our proposed method reduces the WER in the region with highest WER by 3.2\% relative and reduces the overall WER by 1.3\% relative. Moreover, our results suggest that at least in our setting, adapting the language modeling component of the system is important for reducing the performance gap. Our empirical results focus on reducing geographic differences in ASR performance, but our method is equally applicable to other scenarios with a need to adapt a model to a specific dataset without degrading overall performance. 

\vfill
\pagebreak
\balance

\bibliographystyle{IEEEtran}
\bibliography{mybib}
\end{document}